\documentclass[a4paper,twoside,twocolumn,english]{revtex4}
\usepackage{times}
\usepackage[T1]{fontenc}
\usepackage[latin1]{inputenc}
\usepackage{array}
\usepackage{graphicx}

\makeatletter

\providecommand{\tabularnewline}{\\}

\usepackage{babel}
\makeatother
\begin{document}

\title{The Fraunhofer Quantum Computing Portal\\
www.qc.fraunhofer.de{\small }\\
{\normalsize A web-based Simulator of Quantum Computing Processes}}

\author{Helge Rosé, Torsten Aßelmeyer-Maluga, Matthias Kolbe, Falk Niehörster
and Andreas Schramm}

\address{Fraunhofer Institute for Computer Architecture and Software Technology\\
Kekuléstraße 7, D-12489 Berlin, qc@first.fraunhofer.de}

\begin{abstract}
Fraunhofer FIRST develops a computing service and collaborative workspace
providing a convenient tool for simulation and investigation of quantum
algorithms. To broaden the twenty qubit limit of workstation-based
simulations to the next qubit decade we provide a dedicated high memorized
Linux cluster with fast Myrinet interconnection network together with
a adapted parallel simulator engine. This simulation service supplemented
by a collaborative workspace is usable everywhere via web interface
and integrates both hardware and software as collaboration and investigation
platform for the quantum community. The modular design of our simulator
engine enables the application of various implementations and simulation
techniques and is open for extensions motivated by the experience
of the users. The beta test version realizes all common one, two and
three qubit gates, arbitrary one and two bit gates, orthogonal measurements
as well as special gates like Oracle, Modulo function and Quantum
Fourier Transformation. The main focus of our project is the simulation
of experimentally realizations of quantum algorithms which will make
it feasible to understand the differences between real and ideal quantum
devices and open the view for new algorithms and applications. That's
why the simulator also can work with arbitrary Hamiltonians yielding
its unitary transformation, spectrum and eigenvectors. To realize
the various simulation tasks we integrate various implementations.
The test version is able to simulate small quantum circuits and Hamiltonians
exactly, the latter through the use of a standard diagonalization
procedure. Circuits up to thirty qubits can be simulated exactly as
well; Hamiltonians of that size, however, have to be approximated
according to the Trotter formulae. For a restricted gate set we also
develop a tensor-sum implementation, which makes it feasible to investigate
circuits with up to sixty qubits.
\end{abstract}
\maketitle

\section{Introduction}

Classical simulation of quantum processes seams like a Don Quixote
enterprise. To simulate 31 qubits we need 32 GByte of memory, and
every additional qubit will double the required resources: time, memory,
power and space. Even the biggest supercomputers, if we could use
them for quantum gate simulations, will give up around the forty qubit
limit. This not only shows the surrender of the classical simulation,
it also gives an impressive illustration of the exponential power
of a quantum computer. So why should we make an effort in classical
simulation of quantum systems?

There are some reasons, and the most important one may be: to get
the knowledge for building a useful quantum computer today, not tomorrow.
Realized quantum computers are very small (around seven qubits). Our
quantum computing simulator is able to simulate 31 qubits in an easy-usable,
reproducible way without the obstacles of experimental setups. The
quantum computing community need not to wait anymore for the next
generation quantum computer -- new algorithms and ideas can be tested
today.

Another good reason is the possibility of using the quantum computing
simulator in a way like a chip simulator in semiconductor industry:
as an useful tool in the first stage of the design process of new
circuits. With its help it is possible to consider all conceivable
quantum devices without a restriction to the experimental realizable
ones. This opens the view for new ideas and concepts usable to invent
new algorithms.

An important point is that simulation makes it possible to compare
ideal quantum circuits with their experimental realizations. In an
experiment the ideal dynamics of any real quantum device is disturbed
by errors and decoherence. A simulator can describe both the ideal
dynamics and its errors and decoherence effects. The non-ideal effects
can be added in a controllable way which gives us an offer to test
modifications of ideal algorithms improving their feasibility to run
under real world conditions.

Finally, a quantum simulator may also be considered as an educational
tool. Quantum mechanics is very demanding for human conceivability,
yet it is the fundamental key for development and use of quantum computers.
Anything which makes the processes of quantum computing more comprehensible
will promote the development of this new kind of information processing.
The visualization of the quantum computing processes is an important
step to improve its public understanding. But not only the public
image of Quantum Information Processing will get more lucidity, the
simulator shows the computer scientist how quantum waves and particles
process information and it will help the physicist to learn that quantum
mechanics can be used for more then the mere description of the material
world.\\

These arguments illustrate the usefulness of quantum simulations and
lead to the question: What software concept should we use to realize
it? There are two contradicting requirements to regard. At first,
classical simulation of quantum circuits requires high memorized hardware.
A reasonable compromise between qubit size and costs is a multiprocessor
cluster with standard boards. This induces the fact that the implementation
of the simulator has to be parallel, which deeply restricts the group
of potential users. On the other hand, only an easily accessible simulator
without special hardware requirements would be a practicable and useful
tool for the quantum computing community. The contradiction between
hardware requirements and public availability can be solved by the
software concept of an web-based computing service.

This service is accessible from everywhere by everybody with a standard
web-browser. The user draws up a simulation task with help of a small
browser applet and sends this simulation job to a server. This server
can be equipped with all necessary parallel hardware and dedicated
parallel simulator engines. By this concept we get two advantages:
The simulator implementation can be optimally adapted to the hardware
employed, and there is no need to make it compatible with diverse
user platforms. The user is freed from any installation, administration
and support tasks which could be slightly complicated in case of parallel
hardware.\\

The Fraunhofer Quantum Computing Service provides a dedicated 56 GByte
RAM Linux cluster with fast Myrinet interconnection network enabling
simulations up to 31 qubits. We have supplemented the simulator by
a collaborative workspace based on the Plone/Zope \cite{plone} content
management system which opens the possibility not only to simulate
problems but also to exchange, publish or discuss simulation runs,
documents and ideas within the user community. We use a modular software
design for our simulator engine taking into account the diversity
of implementations and simulation techniques needed for the simulation
of quantum processes. Additionally, this modular concept makes the
simulator easily adaptable and open for extensions motivated by the
experience of the users.

The next section will outline structure and rationale of the architecture
of the simulator. Then we will give a short description of the physical
background and numerical techniques used for the quantum simulations
engines. The last section will discuss the potentials and future developments
of the Fraunhofer Quantum Computing Services.

\section{Software Concept and Technology}

\begin{figure}[h]
\includegraphics[%
  width=0.50\textwidth]{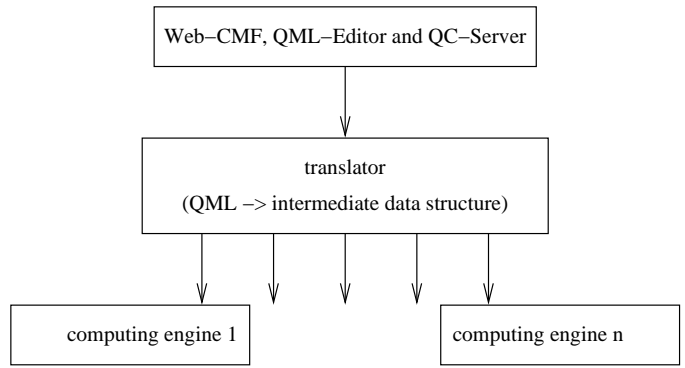}

\caption{\label{cap:Software-architecture-of}Software architecture of the
Fraunhofer Quantum Computing Simulator.}
\end{figure}

The structure of the Fraunhofer Quantum Simulator is given by three
main components: Web Interface, translator, and various modular computing
engines (see Figure~\ref{cap:Software-architecture-of}). 

\begin{itemize}
\item The Web Interface provides graphical editing of quantum gate circuits
and interaction graphs, and handles all the administrative work. It
has been written in Java. 
\item For the various simulation services offered, there is a number of
\textit{computing engines} (see below). 
\item The Web Interface contains a \textit{configurator,} whose principal
activity is to analyze the job submitted by the user so as to infer
the appropriate computing engine, the memory demand, the number of
processors, and the expected computation time. 
\item The Web Interface communicates with the \textit{translator} in both
directions by files with QML texts (\textit{Quantum Markup Language}).
These files describe the jobs and their results, respectively, in
a self-contained fashion. Syntactically, QML is an XML subset and
hence human-readable; for more information see the online documentation. 
\item Apart from the activities in the web interface, the execution of a
job consists of the following steps: 

\begin{enumerate}
\item The translator translates the QML input to a language-inde\-pen\-dent
data structure (see below). 
\item The translator invokes the computing engine selected by the configurator,
passing said data structure. 
\item The computing engine performs the requested computations and writes
the results to an output file. 
\item The translator constructs a QML result file for the web interface
from the output of the computing engine and the original input file. 
\end{enumerate}
\end{itemize}
All in all, the system makes the computing power of a parallel machine
accessible at an ease of use similar to that of a pocket calculator.

\paragraph{Input and output data.}

The data passed from the translator to the computing engine consists
of an operator/operand tree and administrative information. The tree
describes a concatenation of quantum gates as product of the corresponding
operators on the state space. These gates comprise {}``conventional''
ones such as CNOT, Toffoli, etc., as well as measurement gates (probabilistic
projection operators), and finally exponentials of subtrees that represent
sums of Hamiltonians.

As to the output of gate simulations, after every time step (which
may contain several gates as long as they do not act on common bits),
the following information about the state~$s$ is made available:
(i)~The Bloch vectors, i.e., the scalar products $\langle s,\sigma_{\ldots}^{(i)}s\rangle$
for each $\sigma_{x}$, $\sigma_{y}$, and~$\sigma_{z}$ and each
$i^{\textrm{th}}$~quantum bit; (ii)~the amplitudes and phases of
those base states whose amplitude exceeds a certain threshold; (iii)~the
entropy.

\paragraph{The computing engines.}

\begin{table*}[!t]
\begin{tabular}{>{\raggedright}p{0.35\textwidth}>{\raggedright}p{0.60\textwidth}}
\hline 
Purpose&
Computing engine \tabularnewline
\hline
\hline 
simulation of quantum gate circuits and

Hamiltonians&
state representation by distributed tree, actions of small matrices
only; approximation of Hamiltonians by Trotter-Suzuki formulae \cite{Tro:59,Suz:77}\tabularnewline
\hline 
simulation of quantum gate circuits and

Hamiltonians of limited size&
construction, diagonalization, exponentiation, and application of
entire\textasciitilde{}matrix (e.g., by Householder method \cite{Gol:96})\tabularnewline
\hline 
simulation of quantum gates circuits

(no Hamiltonians) of larger size&
space-saving state representation through analysis of gate topology\tabularnewline
\hline 
simulation of quantum gates

(no Hamiltonians) of larger size&
approximated state representation by truncated series of tensor states\tabularnewline
\hline 
computation of full spectrum of a

Hamiltonian of limited size&
construction and diagonalization of entire matrix\tabularnewline
\hline 
computation of margins of spectrum of

a Hamiltonian&
Lanczos method \cite{Gol:96}\tabularnewline
\hline
\end{tabular}

\caption{\label{cap:The-various-computing}The computing engines}
\end{table*}
The actual numerical work is done by the computing engines. Each computing
engine solves a particular problem by a particular algorithm and data
representation with its individual advantages. In other words, the
computing engines populate a two-dimensional space whose dimensions
might be called {}``problem classes'' and {}``solution concepts''.
This set of computing engines is the place where future extensions
are likely to be incorporated. Therefore, in order not to hamper creativity,
no structure is preimposed on this set. Adding a computing engine
essentially involves the following actions: 

\begin{itemize}
\item Programming the computing engine as a C++ class that meets a certain
interface; 
\item devising criteria when to favor this computing engine over the others,
and incorporating these criteria in the configurator.
\end{itemize}
The computing engines that are currently available or under construction
are sketched in Table~\ref{cap:The-various-computing}. They employ
various state representation concepts and algorithms, which, for instance,
exploit certain redundancies of the quantum gate topology or certain
approximations of the state representation in order to make the simulation
of larger systems possible. The details will be fixed and published
later. The parallel implementations use MPI\@.

Flexibility is further facilitated by a the fact that the QML language
can easily be extended with new element and attribute names, and that
the internal intermediate data structure shields the computing engines
from language idiosyncrasies.

\section{Quantum Simulation}

This section describes the physical background of the quantum simulations
engines. The simulation of quantum systems is one of the most complicated
problems in physics. To define this problem, consider a time-dependent
state $\left|\psi(t)\right\rangle $ in some Hilbert space. The dynamics
of this state is given by the Schrödinger equation\begin{equation}
i\frac{\partial}{\partial t}\left|\psi(t)\right\rangle =\widehat{H}\left|\psi(t)\right\rangle \label{tdse}\end{equation}
 with Hamiltonian $\widehat{H}$. Methods to solve this equation emerge
as important tools to simulate for instance molecular \cite{KosKos:83,Hor:84}
and nuclear collisions \cite{FloKooWei:78,GruMueSch:82,Neg:82}, atom-surface
interactions \cite{KosCer:84}, high-resolution electron-microscopy
image simulation \cite{Spe:81}, light propagation in optical fibers
\cite{FeiFle:80}, electron motion in disordered materials \cite{KraMcKWea:81}
etc. A general overview can be found in \cite{DolGub:90} for instance.
The formal solution to equation (\ref{tdse}) is given by\begin{equation}
\left|\psi(t)\right\rangle =\exp(-it\widehat{H})\left|\psi(0)\right\rangle \label{solutionTDSE}\end{equation}
 which is complicate to calculate. First we remark that the Hamiltonian
$\widehat{H}$ is an element of the Lie algebra of the automorphisms
of the Hilbert space, which is the unitary group $U(n)$ for some
natural number $n$ including the case $n\rightarrow\infty$ which
is the dimension of the Hilbert space. Then the exponential is a map
from the Lie algebra to the Lie group. Thus, the generator $\exp(-it\widehat{H})$
of the formal solution generates for all possible Hamiltonians all
unitary transformations, i.e., all elements of the group $U(n)$.
Therefore we obtain two possibilities to generate unitary transformations
which serve as quantum computing operations: by fixing a Hamiltonian
or a unitary matrix. The so-called \char`\"{}Quantum Emulator\char`\"{}
of De Raedt et.al. \cite{DeRHamMicDeR:00} is the most important example
of a quantum computing simulator using a Hamiltonian for the computation.

Now we choose a 2-level system describing the qubit \[
\left|\varphi\right\rangle =a\left|0\right\rangle +b\left|1\right\rangle \;\;\;|a|^{2}+|b|^{2}=1\]
 which can concretely represented by the spin $\left|0\right\rangle ={\uparrow}$
and $\left|1\right\rangle ={\downarrow}$. Next we orient the spin
in the z-direction, i.e. the Pauli matrix\[
\sigma_{z}=\left(\begin{array}{rr}
1 & 0\\
0 & -1\end{array}\right)\]
 is diagonal. The other Pauli matrices are given by\[
\sigma_{x}=\left(\begin{array}{rr}
0 & 1\\
1 & 0\end{array}\right),\qquad\sigma_{y}=\left(\begin{array}{rr}
0 & i\\
\!\!-i & 0\end{array}\right)\:.\]
 Then the Hilbert space for $N$ qubits is a $2^{N}$-dimensional
complex vector space, and a state is the sum of tensor states. In
our simulator we use the two possibilities described above to choose
a unitary transformation. At first there is a library of quantum gates
or unitary gates (CNOT, Toffoli, etc.) represented by matrices, for
instance \[
\left(\begin{array}{cccc}
1 & 0 & 0 & 0\\
0 & 1 & 0 & 0\\
0 & 0 & 0 & 1\\
0 & 0 & 1 & 0\end{array}\right)\]
 is the CNOT gate. Furthermore there is a collection of scalable gates
like QFT, Grover Step, Oracle and the Grover gate. Among them there
is a special gate, the EXP gate which is unitary gate by fixing the
Hamiltonian.

Before we describe the EXP gate, we have to fix some notation. Let
$\overrightarrow{\sigma}=(\sigma_{x},\sigma_{y},\sigma_{z})$ be the
{}``vector'' of Pauli matrices and $\sigma_{x,y,z}^{(i)}$ is the
action of a Pauli matrix on the $i$-th qubit, i.e.\[
\sigma_{x,y,z}^{(i)}=\underbrace{1\otimes1\otimes...\otimes}_{i-1}\sigma_{x,y,z}\otimes\underbrace{1\otimes...\otimes1}_{N-i}\]
 where $1$ is the $2\times2$ unit matrix. For the Hamiltonian, we
assume only nearest neighbor interactions, i.e. only 2-qubit interactions.
That is, we have to fix two couplings matrices $E_{ij}^{(2)},E_{i}^{(1)}$
for the 2-qubits interactions and for the 1-qubit interaction with
an external field, respectively. Furthermore we need the adjacency
matrix $J_{ij}$ containing the interaction structure of the qubits.
Finally we obtain our Hamiltonian\begin{equation}
\widehat{H}=\sum_{i<j}\: J_{ij}\left(\left(\overrightarrow{\sigma}^{(i)}\right)^{T}E_{ij}^{(2)}\overrightarrow{\sigma}^{(j)}\right)+\sum_{i}\: E_{i}^{(1)}\overrightarrow{\sigma}^{(i)}\label{hamilton}\end{equation}
 where $\overrightarrow{\sigma}^{T}$ denotes the transpose Pauli
{}``vector'' with the obvious rule\[
E_{ij}^{(2)}=\left(E_{ji}^{(2)}\right)^{T}\;.\]
 To illustrate this, we write down the ordinary quantum Ising model,
i.e. with $\sigma_{z}\otimes\sigma_{z}$-coupling and the external
field into the $\sigma_{x}$-direction. Then we have to fix the matrices
for all combinations of $i,j$ to be\[
E_{ij}^{(2)}=\left(\begin{array}{ccc}
0 & 0 & 0\\
0 & 0 & 0\\
0 & 0 & E_{0}\end{array}\right)\qquad E_{i}^{(1)}=\left(\begin{array}{c}
B\\
0\\
0\end{array}\right)\]
 with $B$ as the external field and $E_{0}$ as interaction energy. 

Now we are left with only one problem: the exponential of the Hamiltonian
to get the unitary transformation related to the Hamiltonian. For
small values of $N$, the number of qubits, one can calculate the
spectrum of $\widehat{H}$ to express the exponential according to
the rules of linear algebra. But for higher values of $N$ we use
an approximation. One method is the so-called Cranck-Nicholson method
where the approximation is given by\[
\exp\left(-it\widehat{H}\right)\approx\frac{2-it\widehat{H}}{2+it\widehat{H}}\]
 where we have the problem to calculate the inverse $\left(2+it\widehat{H}\right)^{-1}$which
is also complicated enough. Here we choose another method know as
Trotter-Suzuki formula \cite{Tro:59,Suz:77}. Lets assume that the
Hamiltonian is a sum of two terms\[
\widehat{H}=H_{0}+H_{1}\]
 and we obtain from the Trotter-Suzuki formula the approximation

\[
\exp\left(-it\widehat{H}\right)\approx\left(\exp(-itH_{0}/n)\exp(-itH_{1}/n)\right)^{n},\]
which is correct to order $t$. In \cite{deR:87}, De Raedt introduces
a second and a fourth order refinement of this formula. The second
order approximation is given by\begin{eqnarray*}
\exp\left(-it\widehat{H}\right) & \approx & \left(\exp(-itH_{1}/(2n))\exp(-itH_{0}/n)\right.\\
 &  & \left.\exp(-itH_{1}/(2n))\right)^{n}\end{eqnarray*}
 correct of order $t^{3}$. In our simulator we implement this approximation
whereas in a future extension we will also implement the forth order
approximation given by\begin{eqnarray*}
\exp\left(-it\widehat{H}\right) & \approx & \left(\exp(-itH_{0}/(2n))\exp(-itH_{1}/(2n))\right.\\
 &  & \exp(it^{3}C/n^{3})\exp(-itH_{1}/(2n))\\
 &  & \left.\exp(-itH_{0}/(2n))\right)^{n}\end{eqnarray*}
 correct of order $t^{5}$ where the operator $C$ is given by\[
C=[H_{0}+2H_{1},[H_{0},H_{1}]]/24\;.\]
Of course we use these formulas recursively to calculate the full
Hamiltonian (\ref{hamilton}). At the end of this section we will
describe the output of a simulation. In the current version of the
simulator we have 3 kinds of output: the Bloch vector, a kind of entropy
as well as the probability and phase of the most important base vectors.
Lets consider a state \[
\left|\psi\right\rangle =\sum_{i=0}^{2^{N}-1}\; c_{i}\left|\phi_{i}\right\rangle \]
 with base vectors $\left|\phi_{i}\right\rangle .$ Then the probability
$p_{i}$ and the phase $\varphi_{i}$ of a base vector is given by
\[
p_{i}=|c_{i}|^{2}\qquad\varphi_{i}=\arg(c_{i})\]
 and we define the entropy $S(\left|\psi\right\rangle )$ of $\left|\psi\right\rangle $
to be\[
S(\left|\psi\right\rangle )=-\sum_{i}|c_{i}|^{2}\frac{\ln(|c_{i}|^{2})}{\ln(2)}\;.\]
\begin{figure*}
\includegraphics[%
  width=0.95\textwidth]{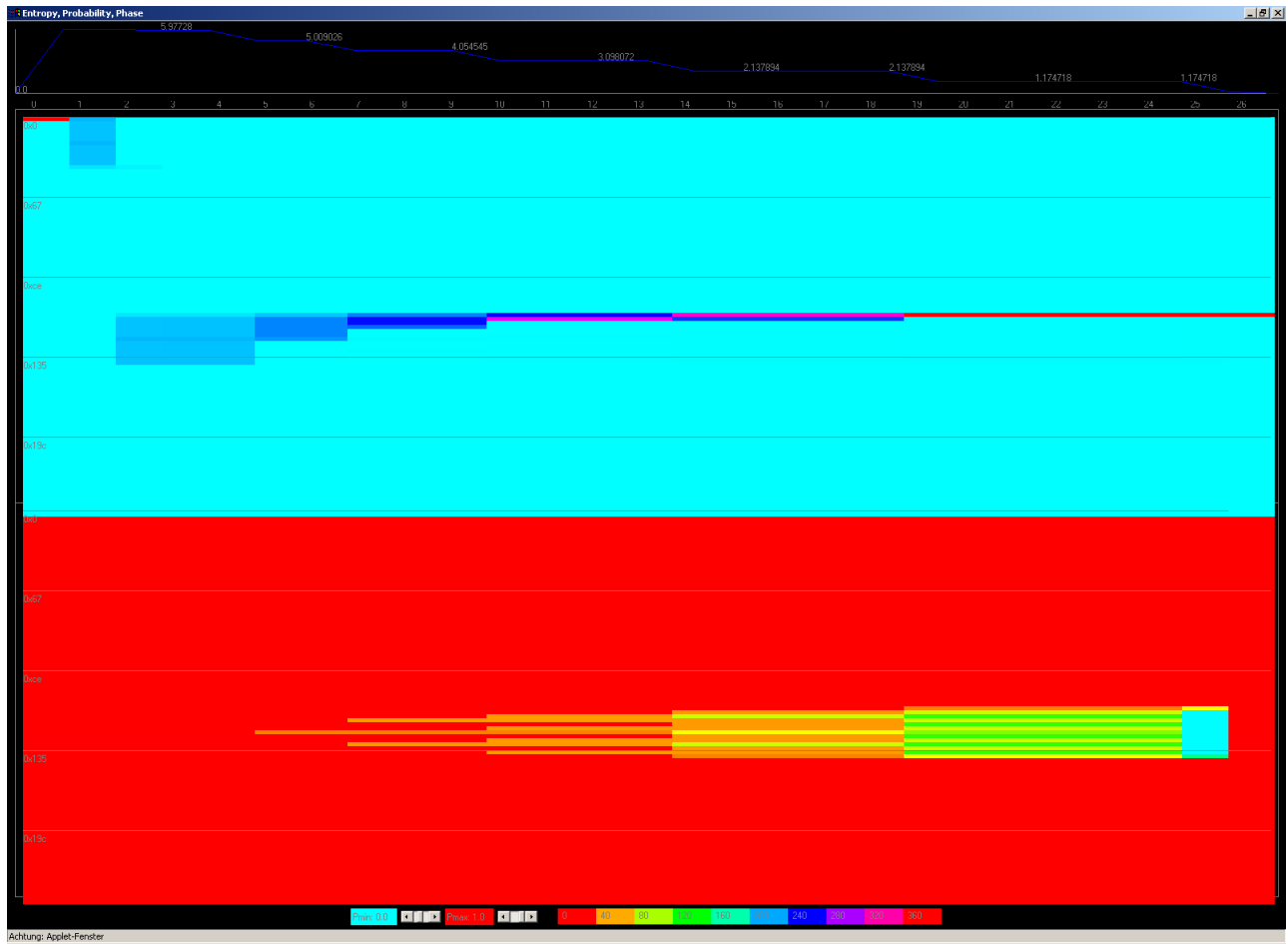}

\caption{\label{cap:Shor-algorithm}Simulation result for a 9 qubit Shor algorithm.}
\end{figure*}
The Bloch vector $\overrightarrow{v_{i}}$ for the $i$-th qubit is
given by the expectation value\[
\overrightarrow{v_{i}}=\left\langle \psi\right|\overrightarrow{\sigma}^{(i)}\left|\psi\right\rangle \]
 in the notation above.

Before we close this section we will remark that the Hamiltonian above
will be extended to more realistic cases like NMR by adding a periodic
term in the matrix $E_{i}^{(1)}$. Furthermore we remark that the
structure of the Hamiltonian above includes also all interesting cases
known from condensed matter physics. There, the electron creation
$a^{+}$ and annihilation $a$ operators are needed. With the settings
$a^{+}=(\sigma_{x}-i\sigma_{y})/2$ and $a=(\sigma_{x}+i\sigma_{y})/2$
one can formulate a substitute of the model in terms of the Hamiltonian
(\ref{hamilton}).

\subsection*{Example: Shor algorithm}

Here we will describe the output of the Shor algorithm for the number
$N=899$ with the random number $a=11$ (see Figure \ref{cap:Shor-algorithm}).
The starting point is the division of all qubits into two registers:
the $x$-register $\left|x\right\rangle $ and the $y$-register $\left|y\right\rangle $
where the length $n_{x}$ of the $x$-register is at least twice the
length $n_{y}=n_{x}/2$ of the $y$-register. In this case we need
$n_{y}=10$ qubits for the $y$-register encoding the number $N=899$
and at least $n_{x}=20$ qubits for the $x$-register giving 30 qubits
for the whole circuit. Here we use the full capacity of the simulator
with 31 qubits. At first (time step 1) we construct a Hadamard state
on the $x$-register, i.e.\[
\left|x\right\rangle =\bigotimes_{i=1}^{n_{x}}\frac{1}{\sqrt{2}}\left(\left|0\right\rangle +\left|1\right\rangle \right)\]
 Then we have the apply the MODULO Gate (time step 2)\[
\left|x\right\rangle \otimes\left|y\right\rangle \Rightarrow\left|x\right\rangle \otimes\left|(y\oplus(a^{x}\bmod N))\right\rangle \]
 with a measurement (time step 3) of the $y$-register afterwards.
Now we have an entanglement between both registers, as can be seen
by the vanishing of some Bloch vectors. Beginning with time step 4,
the quantum Fourier transformation {}``removes'' all unnecessary
states by interference and we are left with one state with high probability
and a lot of states with low probability. Thus after the measurement
of the $x$~register the algorithm is finished. A short look into
the Bloch vector of the $x$-register shows the value $M=954733$.
A continuous fraction expansion leads to the fraction represented
by $[10,5,1,3,9,1,6,3]$ and we obtain for the proposed value of the
exponent $r=7$. Only factors of this value are the solution and in
our case \[
11^{210}\bmod899=869\]
Finally we are looking for the greatest common divisors (gcd)\[
gcd(869-1,899)=31\]
\[
gcd(869+1,899)=29\]
and we obtain the factors $29$ and $31$ of the number $899$.

The Figure \ref{cap:Shor-algorithm} visualizes the main base states
and their phases as well as the entropy of the states of a 9 qubit
Shor algorithm with a 6 qubit x~register and a 3 qubit y~register.
The problem is the factorization of $N=6$ with a random number $a=4$.
We have done the simulation also for a 31 qubit Shor algorithm. The
results look similar, but possess a huge number of base states, which
are too many for a instructive visualization. After the first time
step the entropy increases to the size of the x~register, and we
obtain all possible base states. Then the modulo gate mixes these
states with the states of y~register, and the measurement thins out
some of them. Beginning with time step~4, the most interesting part
starts -- the quantum Fourier transformation. During this process
the number of interesting base states decreases and we finish with
only one state with a high probability and many states of low probability.

\section{Quantum Investigations at the Internet~Cafe}

We have seen that the Fraunhofer Quantum Computing Service is characterized
by three decisive aspects: it is a web-based computing service, it
combines a simulator with a collaborative workspace, and it has a
modular, extensible, and open software design. In the following let
us discuss each aspect and its potentials for future developments.

The web-based design gives the simulator a ubiquitous availability.
There is no need to install any software or to buy special hardware,
the computing service makes quantum investigations feasible everywhere
-- at the institute, at home or even at the Internet cafe. All simulation
runs are stored at the server and can be accessed everywhere and anytime.
Especially this advantage could be a source of mistrust: the own ideas
and investigations should stored on a foreign server? This sensitive
aspect has to be justified by strong security protocols and high privacy
standards. Each simulation job has to be encoded and unambiguously
signed by owner and date making the authorship pursuable.

Closely connected to this point is the collaborative aspect of the
portal. The central availability, publishing and discussion of the
newest research data and documents is a key element of scientific
success. The integrated content management system of the portal supports
all appropriate forms of collaborative work like publishing with and
without review, discussion groups, news, wiki systems etc. The quantum
circuits and simulation results can be exchange by the uniform QML
document format.

An important feature of the QML-based data definition is its hypertext
ability. QML files with well-tried quantum circuits can be stored
on any web server building a library of established modules of quantum
algorithms. This modules can be included as sub-circuits in other
QML files by a simple HTML reference. This facilitates the self-organized
formation of a public library of quantum algorithms by the community.

This community-based building of a library of quantum circuits can
be supplemented by an establishment of a network of simulation servers.
The modular design of the simulation engines provides an easy way
for integration of already used simulation software developed by other
research groups. The requirements to integrate other simulators in
our system are very low. The software has to be extended by a QML
parser which will realize a common interface for all various simulators.
Then, the new QML-capable engine could be run at our cluster providing
a new computing service for the community. But it is also possible
to install a small Java-server on the cluster of the partner group
dispatching all simulation jobs which requesting the new engine. The
user will not see this infrastructure details, he will only note that
the portal offers more and more simulation features which are provided
by more and more research groups of the community. Currently we start
a cooperation with the quantum computing group at the University of
Newcastle to integrate group-theoretical quantum algorithms.

We hope that our simulator and workspace will become a useful tool
for the community and we will get many user responses and comments
which help us to stabilize, improve and extent the system leading
to the one objective of our intension: to support all efforts which
brings the quantum computer from dream to reality.

\section{Acknowledgement}

This work is supported by the German Federal Ministry of Education
and Research, and we thank Marius van der Meer for facilitating the
EIQU Project (FZK: 01IBB01A). We thank the group of Thomas Beth, especially
Dominik Janzing, for fruitful discussions, and Uwe Der for his hard
effort and deep support to bring the cluster running. Our special
thank goes to Stefan Jähnichen -- without whose great dedication this
project would never have been started.

\bibliographystyle{apsrev}
\bibliography{eiqu-intro}

\end{document}